\documentclass{SciPost}

\hypersetup{
    colorlinks,
    linkcolor={red!50!black},
    citecolor={blue!50!black},
    urlcolor={blue!80!black}
}

\usepackage[bitstream-charter]{mathdesign}
\DeclareSymbolFont{usualmathcal}{OMS}{cmsy}{m}{n}
\DeclareSymbolFontAlphabet{\mathcal}{usualmathcal}

\fancypagestyle{SPstyle}{
\fancyhf{}
\lhead{\colorbox{scipostblue}{\bf \color{white} ~SciPost Physics }}
\rhead{{\bf \color{scipostdeepblue} ~Submission }}

\fancyfoot[C]{\textbf{\thepage}}
}

\usepackage{bm}
\usepackage{braket}
\usepackage[normalem]{ulem}

\newcommand{\mycomment}[1]{}

\begin{document}

\pagestyle{SPstyle}

\begin{center}{\Large \textbf{\color{scipostdeepblue}{
Tetrahedral Core in a Sea of Competing Magnetic Phases in Graphene
}}}
\end{center}

\begin{center}\textbf{
\newcommand{\surname}[1]{#1}
Maxime \surname{Lucas}\textsuperscript{1$\star$},
Arnaud \surname{Ralko}\textsuperscript{2$\dagger$},
Andreas \surname{Honecker}\textsuperscript{1}, \\ and
Guy \surname{Trambly de Laissardi\`ere}\textsuperscript{1$\ddagger$}
}
\end{center}

\begin{center}
{\bf 1} Laboratoire de Physique Th\'eorique et Mod\'elisation,
CY Cergy Paris Universit\'e, \\ CNRS UMR8089,
95302 Cergy-Pontoise, France
\\
{\bf 2} Institut Néel, Université Grenoble Alpes, CNRS, UPR2940, Grenoble INP, 38042 Grenoble, France
\\[\baselineskip]
$\star$ \href{mailto:maxime.lucas-guerreau@cyu.fr}{\small maxime.lucas-guerreau@cyu.fr}\,,\quad
$\dagger$ \href{mailto:arnaud.ralko@neel.cnrs.fr}{\small arnaud.ralko@neel.cnrs.fr}\,,\quad
$\ddagger$ \href{mailto:guy.trambly@cyu.fr}{\small guy.trambly@cyu.fr}
\end{center}

\section*{\color{scipostdeepblue}{Abstract}}
\textbf{\boldmath{
We demonstrate the emergence of a robust tetrahedral magnetic ground state in monolayer graphene doped to the van Hove singularity (vHS). This noncoplanar, gapped spin configuration—featuring four equally inclined moments—has been previously identified as a candidate instability.
Here, not only do we confirm its stability across all finite interactions using fully self-consistent, real-space-resolved calculations, but we also go beyond earlier work by charting the full surrounding phase diagram. In doing so, we unravel a cascade of symmetry-broken magnetic states — pseudo-tetrahedral, planar, collinear, and modulated textures — which we classify using spin structure factors and vector order parameters.
These results stem from unrestricted Hartree-Fock simulations on large supercells with dense $k$-point sampling, enabling us to resolve interaction-driven magnetic and charge inhomogeneities.
Our findings connect directly with recent ARPES and doping experiments near the vHS in graphene, and establish the tetrahedral state as the central correlated instability in this regime, offering predictive insight into emergent magnetism in correlated Dirac materials.
}}

\vspace{\baselineskip}

\noindent\textcolor{white!90!black}{%
\fbox{\parbox{0.975\linewidth}{%
\textcolor{white!40!black}{\begin{tabular}{lr}%
  \begin{minipage}{0.6\textwidth}%
    {\small Copyright attribution to authors. \newline
    This work is a submission to SciPost Physics. \newline
    License information to appear upon publication. \newline
    Publication information to appear upon publication.}
  \end{minipage} & \begin{minipage}{0.4\textwidth}
    {\small Received Date: {\today} \newline Accepted Date \newline Published Date}%
  \end{minipage}
\end{tabular}}
}}
}

\vspace{10pt}
\noindent\rule{\textwidth}{1pt}
\tableofcontents
\noindent\rule{\textwidth}{1pt}
\vspace{10pt}

\section{Introduction}
\label{sec:intro}

The exploration of strongly correlated phases in two-dimensional electron systems has been revitalized by the discovery of superconductivity, Mott insulating states, and anomalous Hall effects in twisted bilayer graphene (TBG) near the 
``magic angle''~\cite{Cao18a,Cao18b,Sherkunov18,Guerci22,Kwan25}. In this regime, the moir\'e superlattice gives rise to ultra-flat electronic bands~\cite{trambly10,SuarezMorell10,Bistritzer10,review2526}, strongly enhancing the role of electron-electron interactions and promoting exotic quantum states. A key unifying feature across these systems is the proximity of the Fermi level to van Hove singularities (vHS), where the density of states diverges and interaction effects are dramatically 
enhanced~\cite{GonzalezArraga17,Liu19,Choi19,Xie19,Jiang19,Sharpe19,Klebl19,Xie20,Mesple21,Choi21,Vahedi21_tBLG,Wagner22,review2526}.
Understanding the role of vHS in driving correlated phases is thus essential not only for TBG but also for untwisted monolayer and bilayer graphene systems~\cite{McChesney10,Li12,Wang12,Nandkishore12,Jiang15,Link19,Rosenzweig19,Wan23,Jiang14supercond,Ying18}.

In monolayer graphene, accessing the vHS via doping remains experimentally challenging but feasible through alkali or rare-earth intercalation. Techniques involving K, Ca, Gd, and Yb atoms have allowed systematic tuning of the Fermi level and direct ARPES observation of the Dirac cone's shift~\cite{McChesney10,Link19,Rosenzweig19}. These studies reveal signatures of extended vHS structures induced by electronic interactions and hybridization. More recently, doping-controlled superconducting domes have been observed in bilayer graphene~\cite{Wan23}, drawing intriguing parallels with the superconducting phases in TBG.

From a theoretical perspective, numerous studies have proposed that the vHS regime in graphene—regardless of stacking—hosts a rich competition between magnetism and superconductivity. Functional renormalization group
and density matrix renormalization group
approaches have consistently predicted a chiral spin-density wave,
also referred to as a tetrahedral magnetic order (Tetra), at quarter
filling~\cite{Li12,Wang12,Jiang14supercond,Ying18,Wilhelm2023,Scholle25}. 
However none of these references seem to tackle low interaction values, and some (such as \cite{Jiang14supercond}) cannot distinguish the Tetra order from a spin-charge-Chern liquid.
This phase also competes with $d+id$ superconductivity~\cite{Nandkishore12}, and may also give rise to spontaneous quantum anomalous Hall effects~\cite{Li12,Jiang15}. 
These findings are further corroborated by variational Monte Carlo simulations~\cite{Ying18} and extended Hubbard models~\cite{Jiang15}, which show robustness of the tetrahedral state against next-nearest-neighbor interactions and suggest a universal tendency toward noncoplanar magnetism near the vHS.

The relevance of noncoplanar and geometrically nontrivial spin states extends beyond graphene. Recent works in frustrated magnets report tetrahedral and cuboctahedral orders stabilized by itinerant electrons or competing exchanges. In triangular metals such as Co$_{1/3}$TaS$_2$, tetrahedral triple-$\bm{Q}$ order is linked to large spontaneous Hall conductivities~\cite{Park23}, while cuboctahedral states appear in kagome antiferromagnets~\cite{Messio12,Messio13,lugan_schwinger_2022}. Spin-1 bilinear-biquadratic models on the honeycomb lattice also host eightfold degenerate chiral spin liquids and multipolar orders with non-zero scalar chirality, from the interplay of dipolar and quadrupolar moments~\cite{Pohle23,Pohle24}. These developments underscore broad interest in unconventional textures across lattices and models, motivating their study in itinerant systems like doped graphene.

\begin{figure*}[t!]
\centering
\includegraphics[width=0.8\textwidth]{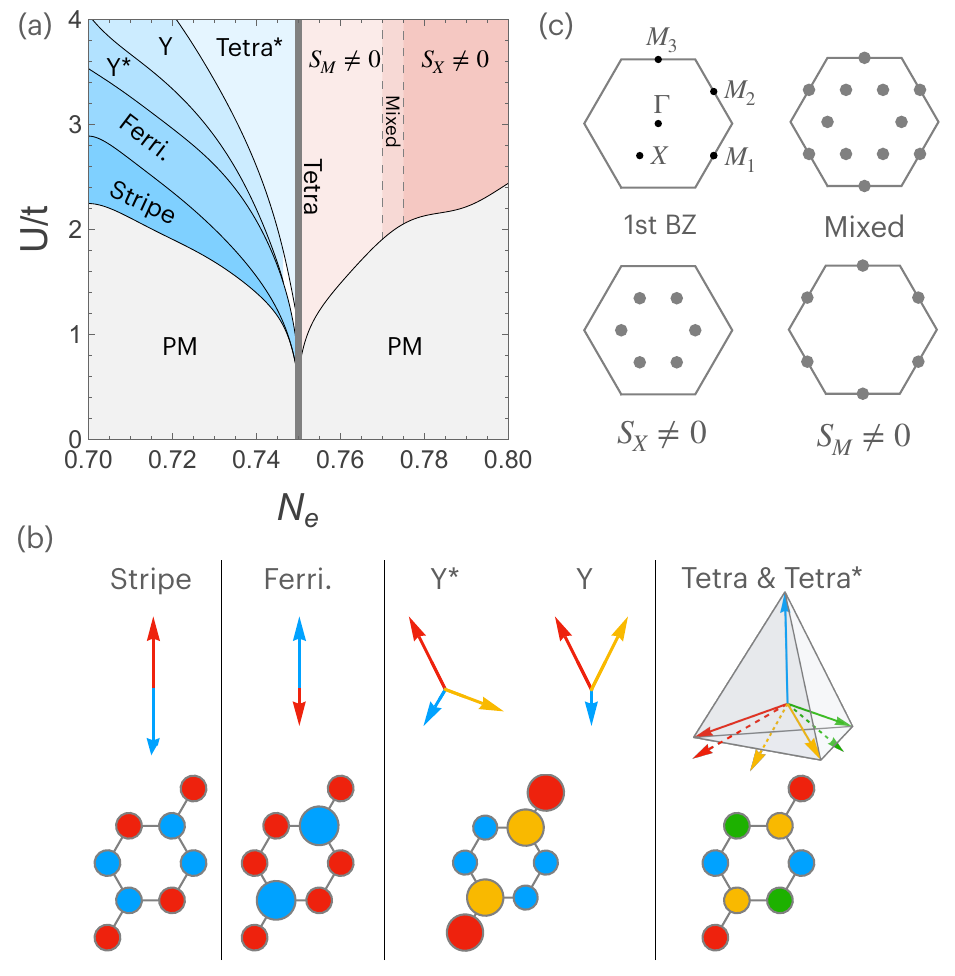}
\caption{\label{Fig_Phase_Diag} 
Magnetic ground-state phase diagram of graphene near quarter doping. (a) Schematic mean-field $(N_e, U)$ phase diagram based on computations at $k_B T = 10^{-7}t$, using a $6\times6$ supercell and $48\times48$ $k$-point sampling.
(b) For doping levels below the van Hove singularity (vHS), $N_e \le 0.75$, indicated by blue regions, all magnetic phases exhibit well-defined spin orders compatible with a $2\times2$ graphene supercell (see also
Fig.\,\ref{fig:2D_proj_6x6}). 
Real-space sketches illustrate these phases: ``Tetra'' denotes the ideal tetrahedral configuration with four spins forming a perfect tetrahedron (found only at $N_e = 0.75$); canted tetrahedral states are its distorted variants and denoted by ``Tetra$^*$''. ``Y'' and ``Y$^*$'' refer to planar ferrimagnetic states with three distinct spin orientations, while ``Ferri.''\ indicates a two-moment ferrimagnetic phase. ``Stripe''\ represents a collinear magnetic phase with uniform spin magnitudes. (c) Above the vHS (red regions), more complex magnetic orders arise, strongly dependent on both doping and Coulomb interaction $U$. These fall into two primary regimes characterized by dominant magnetic structure factors at the high-symmetry points $M_i$ and $X$ in the Brillouin zone. An intermediate mixed region, where both $S_M \ne 0$ and $S_X \ne 0$, is indicated with dashed lines. Phase boundaries are defined according to the order parameters described in Fig.\,\ref{Fig_Order_parameters}.}
\end{figure*}

Here we study magnetic phases in monolayer graphene near the vHS using the Hubbard model within a non-collinear Hartree-Fock framework, resolving the interplay between filling $N_e$ (average number of electron per orbital) and on-site Coulomb interaction $U$ on large supercells with full reciprocal-space resolution. 
Building on mean-field validations at half-filling ($N_e=1$) in graphene and twisted bilayer graphene (e.g., Refs.~\cite{Sorella1992,Feldner10,Marcin19,Vahedi21_tBLG}), we construct an ultra-low-temperature density-interaction $(N_e,U)$
phase diagram\footnote{This holds for all $U/t \le 3.75$ studied; behavior at larger $U$ is beyond this work.}, focusing on quarter doping ($N_e = 3/4 = 0.75$ or $5/4 = 1.25$) where correlation effects peak. 
Figure~\ref{Fig_Phase_Diag} shows that at $N_e=0.75$ we find a \textit{tetrahedral magnetic order} (Tetra)---a noncoplanar, three-dimensional configuration with exact tetrahedral symmetry. This state is gapped and exhibits zero net magnetization despite long-range order. 
Away from quarter filling and for $U \le 3.75t$, the magnetic landscape becomes richer: for $N_e < 0.75$ we identify phases (pseudo-tetrahedral (Tetra$^*$), planar Y and Y$^*$, ferrimagnetic, stripe) robust to finite-size effects, while for $N_e > 0.75$ we observe intricate, often incommensurate textures with emergent \textit{charge density displacements}, possibly linked to local phase separation or magneto-electric coupling. 
These results highlight doping-driven complex spin orders in graphene and support the view that correlated phases---beyond conventional spin-density waves---play a central role near the vHS.

\section{Model and method}
We consider monolayer graphene described by the single-band Hubbard model on the honeycomb lattice
(see, e.g., Ref.~\cite{Yazyev2010}),
\begin{equation}
H = -t\sum_{\langle i,j \rangle,\sigma} (c^\dagger_{i\sigma}c_{j\sigma} + \mathrm{h.c.}) + U\sum_i n_{i\uparrow}n_{i\downarrow} - \mu \sum_{i\sigma} n_{i\sigma},
\label{Eq_Hubbard}
\end{equation}
where $t$ is the nearest-neighbor hopping amplitude, $U$ the on-site Coulomb repulsion, $\mu$ the chemical potential and $c_{i\sigma}$ ($c^\dagger_{i\sigma}$) annihilates (creates) a spin-$\sigma$ electron on the $p_z$ orbital at site $i$.

To investigate magnetic instabilities at finite doping, we solve this model using an unrestricted Hartree-Fock mean-field decoupling \cite{Thu_these,Scholle23Square}. 
The interaction term is then given by
\begin{align}
H_U^{\mathrm{MFT}} = U  \sum_i & \Big[ \langle n_{i\downarrow} \rangle n_{i\uparrow} + \langle n_{i\uparrow} \rangle n_{i\downarrow} - \langle n_{i\uparrow} \rangle \langle n_{i\downarrow} \rangle
- \langle S_i^- \rangle S_i^+ - \langle S_i^+ \rangle S_i^- + \langle S_i^+ \rangle \langle S_i^- \rangle \Big],
\label{Eq_H_MFT}
\end{align}
where $S_i^+ = c^\dagger_{i\uparrow} c_{i\downarrow}$ and $S_i^- = c^\dagger_{i\downarrow} c_{i\uparrow}$.
Most previous investigations concentrated on $N_e=1$
where one may neglect the $S^\pm$ expectation values and simplify to a Hartree approximation
(see, e.g., Refs.~\cite{Yazyev2010,Feldner10,Marcin19,Vahedi21_tBLG,review2526}).
Inclusion of these mean fields allows us to capture arbitrary spin textures
that we find to be essential for a description of the behavior close to the vHS.
We note that the case $U=3t$ has recently been investigated at finite temperature~\cite{Scholle25}.

The local magnetization $\bm{M}_i = (M_x^i, M_y^i, M_z^i)$ and local density $N_e^i$ are defined as, \mycomment{extracted self-consistently from the eigenstates at low temperature}
\begin{equation}
    \begin{split}
        M_x^i=&\frac{\braket{S_{i}^{+}}+\braket{S_{i}^{-}}}{2}, \quad M_y^i=\frac{\braket{S_{i}^{+}}-\braket{S_{i}^{-}}}{2i}, \\
        M_z^i=&\frac{\braket{n_{i\uparrow}}-\braket{n_{i\downarrow}}}{2}, \quad N_e^i = \braket{n_{i\uparrow}}+\braket{n_{i\downarrow}},
    \end{split}
\label{Eq_Mag}
\end{equation}
which lead to a rotational symmetry of the solutions space (SU(2) spin-rotation invariance). While each magnetic ground state would not preserve spin-rotational symmetry, solutions found for the same initial parameters may differ from one another by a rotation and still be equivalent.

We solve the self-consistency conditions by iteration.
Using the initial densities at every site $i$ of the $L\times L$ supercell 
$\left(\braket{n_{i\downarrow}}, \braket{n_{i\downarrow}}, \braket{S_{i}^{+}}, \braket{S_{i}^{-}} \right)$,
the $2\mathcal{N}$$\times$$2\mathcal{N}$ matrix Hamiltonian (Eq.\,(\ref{Eq_Hubbard}) with (\ref{Eq_H_MFT}), the factor $2$ is due to the spin)
is written in the basis of the $\mathcal{N}$ $\rm p_z$ orbitals, $\mathcal{N}=2L^2$, 
and it is diagonalized in reciprocal space to obtain eigenvectors $\Psi_{n}(\bm{k})$ and eigenvalues $E_n(\bm{k})$. 
We then compute  new densities to be used as input for the next iteration by defining the spinor $\Phi = [c_{i,\uparrow} , c_{i,\downarrow}]^T$ such that our mean-field parameters are defined by the matrix $\mathcal{M}$ as 
$\langle \Phi^\dagger \mathcal{M} \Phi \rangle$
whose elements are given by,
\begin{eqnarray}
\mathcal{M}_{\alpha \beta} = \sum_{\bm{k},n} \psi_{n,i,\alpha}^*(\bm{k}) \, \psi_{n,i,\beta}(\bm{k})\, n_{\textrm{FD}}(E_n(\bm{k}),\mu), 
\end{eqnarray}
where 
$\psi_{n,i} = \braket{i\uparrow|\Psi_{n}(\bm{k})}$ and 
$\psi_{n,i+\mathcal{N}} = \braket{i\downarrow|\Psi_{n}(\bm{k})}$, 
$i=1,..., \mathcal{N}$,
are the components of  $\Psi_{n}(\bm{k})$ in the $\rm p_z$ orbitals basis with spin $\uparrow$ and $\downarrow$.
$\bm k$ are the momenta on the grid $N_k$ in the unit cell of the reciprocal lattice corresponding to the $L \times L$ supercell ($L=6$ and $L=24$, {\it i.e.}, $\mathcal{N}=72$ and $\mathcal{N}=1152$ sites in a supercell, respectively).
$n_\textrm{FD}$ is the
Fermi-Dirac function,
$n_\textrm{FD} = {1}/{\left(1+{\rm e}^{(E_n(\bm{k})-\mu)/k_BT}\right)}$
at a non-zero temperature $T$.
The new densities
are compared with the previous densities, and the process is repeated until their difference is less than a certain threshold (usually going below $10^{-7}$ does not change the results). 
Results presented here
 have been obtained for a finite but ultra-low-temperature $k_B T=10^{-7} t$. While the Mermin-Wagner theorem \cite{MW1966} prevents true breaking of a continuous symmetry at finite temperature in an infinite two-dimensional system, we trust that our chosen temperature value is low enough to reach zero-temperature results for all practical purposes. We introduced this ultra-low-temperature consideration as a technical trick to circumvent the usual oscillations found during convergence in zero-temperature self-consistent computations.
For each iteration the chemical potential $\mu$ is calculated by a bisection search around the Fermi energy.  
Equation~(\ref{Eq_Mag}) yields the final magnetic state of the system. 

This approach also enables the construction of a $( N_e,U)$ phase diagram at low temperature, capturing both commensurate magnetic phases and interaction-induced charge inhomogeneities, including nontrivial textures beyond simple spin-density waves. Further numerical details and convergence tests are provided in Appendix \ref{appendixC} where we tackle the importance of a precise $k$-grid, especially for low $U$. We also establish that even with many $k$-points $N_k$, it is possible to get stuck in metastable states which proves the need for repeated computations starting from different initial states.

\begin{figure*}[t!]
\centering
\includegraphics[width=0.8\textwidth]{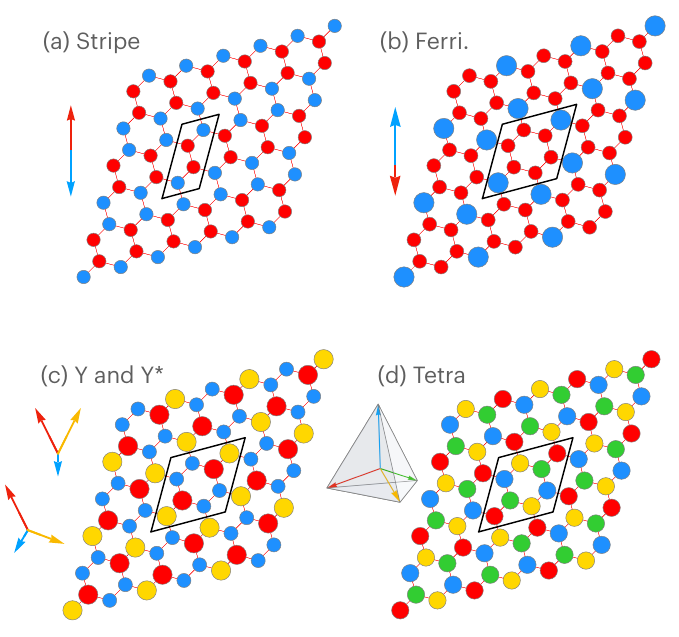}
\caption{
{Real-space magnetic configurations for $N_e \le 0.75$.}
Two-dimensional snapshots of the main magnetic patterns were obtained in the $6 \times 6$ supercell.
Each colored disk represents a lattice site, where the {color} encodes the local spin {orientation} (angle), 
and the {disk size} is proportional to the magnitude of the local magnetic moment.
Panels (a)–(d) correspond respectively to the {stripe} ($U=2.58t$, $N_e=0.71$), {ferrimagnetic} ($U=2.83t$, $N_e=0.72$), {Y/Y$^{*}$} ($U=3t$, $N_e=0.73$), and {tetrahedral/tetrahedral$^{*}$} ($U=3t$, $N_e=0.75$) states. Each configuration is well stabilized at the chosen $U$ and $N_e$ values, but represent their respective phases.
The black outline marks the smallest supercell compatible with each magnetic order.
These maps visualize the spin textures in real space and correspond to the spin configurations shown in each panel.
}
\label{fig:2D_proj_6x6}
\end{figure*}

\section{Phase diagram}
Figure~\ref{Fig_Phase_Diag}(a) presents a sketch of the $(N_e,U)$ phase diagram computed at low temperature for $0.7 \le N_e \le 0.8$ and $U/t \le 3.75$, using a $6\times6$ graphene supercell with $48\times48$ $k$-point sampling (see Appendices \ref{appendixC}, \ref{appendixD} for technical details). This captures the evolution of magnetic ground states near quarter doping, $N_e=0.75$, where the Fermi level crosses the vHS.
Owing to particle-hole symmetry, the same results apply to
$N_e \in [1.2,1.3]$.

At exact quarter doping ($N_e = 0.75$ or $1.25$), the critical interaction $U_c$ vanishes due to the divergent density of states at the vHS. In this limit, the system always stabilizes (with enough $k$-points, see
Appendix \ref{appendixC} for more details) into a tetrahedral magnetic order [Fig.~\ref{Fig_Phase_Diag}(b)], also known under different names in previous investigations~\cite{Li12,Jiang14supercond,Wang12,Ying18,Nandkishore2012,Wilhelm2023,Scholle25}. This noncoplanar configuration spans a $2\times2$ graphene supercell and consists of eight local moments pointing along the directions of a perfect tetrahedron ($109.47^\circ$ angles). It is the only gapped state in the phase diagram and carries zero net magnetization $\bm{M}_t=\bm{0}$. We also wish to highlight that as the tetrahedral order has a non-zero gap, it is present in a finite range of the chemical potential. With respect to average density per site $N_e$ it is indeed only present at one specific value, but is not restricted to one singular value of the chemical potential.

Away from the vHS, magnetism sets in above a finite $U_c$ that grows as $N_e$ deviates from 0.75, consistent with the Stoner criterion $U_c \propto 1/D(E)$, with $D(E)$ the density of states. A paramagnetic phase thus appears at low $U$ (gray). To characterize the surrounding phases, it is useful to distinguish between the regions below and above $N_e = 0.75$.

For $N_e < 0.75$, we identify five well-defined magnetic states,
confirmed by larger supercell ($24 \times 24$) calculations.
Figure~\ref{fig:2D_proj_6x6} may help to illustrate how the orders sketched in Fig.~\ref{Fig_Phase_Diag}
tile the honeycomb lattice.
As $N_e$ decreases, these phases evolve smoothly from the tetrahedral state as:
(i) a Tetra$^*$ state that is a deformation of the tetrahedral state, {\it i.e.}, this state still has moments pointing in four different directions, but these no longer 
are at exactly $109.47^\circ$ angles, and the lengths also become different. Moreover when looking at its band structure, the gap found in the exact Tetra state is still present, while the Fermi energy is shifted outside of said gap; 
(ii) a region where only three different magnetic moments appear,
two of these moments are arranged symmetrically around the third one.
Such a state is reminiscent of the ``Y'' state that is discussed in the context
of the Heisenberg antiferromagnet (see Refs.~\cite{Gvozdikova2011,Gallegos2025} and references therein), and therefore
we use the same notation also here; 
(iii) going even further away from $N_e=0.75$, the Y state distorts with three distinct magnetic moments,  all of them being now nonequivalent.
In analogy with the Tetra$^*$ state, we call this the ``Y$^*$'' state; 
(iv) this Y$^*$ state is then reduced to two distinct moments parallel to each other (collinear),
but with different moment magnitudes. This phase is ``ferrimagnetic'' because of the overall net magnetic moment present in the system; 
(v) eventually, the moment imbalance of the ferrimagnetic state vanishes letting a ``stripe'' order state to be stabilized, 
that is reminiscent of the one reported in the Heisenberg-Kitaev model on the honeycomb lattice \cite{Chaloupka2013,Oitmaa2015}.
This state could be described by a reduced $2 \times 1$ supercell, compare Fig.~\ref{fig:2D_proj_6x6}(a),
while all other states in the region $N_e \le 0.75$ require a $2\times2$ supercell (Fig.~\ref{fig:2D_proj_6x6}(b--d)).
All states except the stripe and pure tetrahedral one
carry a finite total magnetization.
Note that small modulations of the charge density occur in the region $N_e < 0.75$, but these are commensurate with the magnetic order and thus not carrying additional information.

\begin{figure}[t!]
\begin{center}
\includegraphics[width=0.6\textwidth]{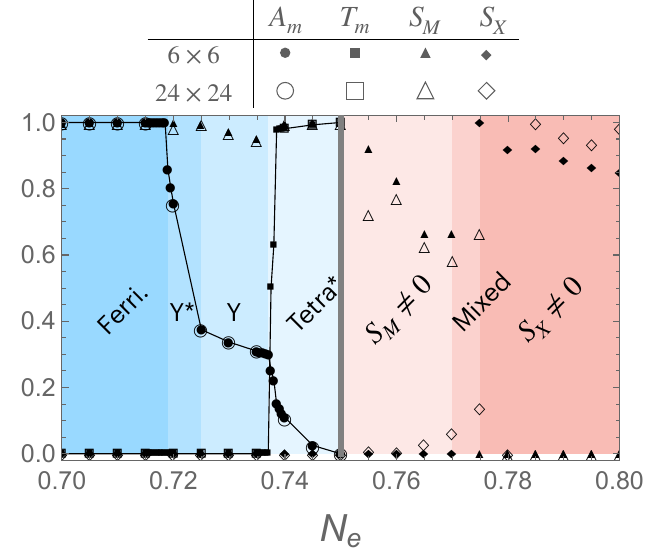}
\caption{\label{Fig_Order_parameters} 
Evolution of magnetic order parameters with electron density in the ground state at fixed interaction strength \( U = 3t \). Upper table: Symbol correspondence with the four observables \( A_m \), \( T_m \), \( S_M \), and \( S_X \) described in the text.
Background colors correspond to those used characterizing the regions of the phase diagram
Fig.\,\ref{Fig_Phase_Diag}. Empty (filled) symbols represent data obtained on \( 6 \times 6 \) (\( 24 \times 24 \)) supercells. Each order parameter is normalized such that they are divided by their maximum value ($A_m=A_m/\max(A_m)$ for example).
}
\end{center}	
\end{figure}

For $N_e > 0.75$, the phase diagram becomes more intricate. As shown in Fig.~\ref{Fig_Phase_Diag}(c), we distinguish two regimes based on the spin structure factor for a $6 \times 6$ supercell.
In the region labeled
$S_M \ne 0$, 
magnetic orders remain quasi-$2\times2$ periodic, with increasingly distorted tetrahedral-like motifs. Beyond $N_e \approx 0.775$, the system enters a region
labeled $S_X \ne 0$, 
where complex $3\times3$ periodic structures emerge, involving over ten nonequivalent spin orientations. The boundary between these two regions
is not sharply defined, and several intermediate states display features of both (see next section and Fig.\,\ref{Fig_Order_parameters}). While the $6\times6$ results give clear trends, full characterization of the
region $S_X \ne 0$
likely requires larger supercells and additional order parameters. Our $24\times24$ simulations confirm the trend toward $2\times2$ and $3\times3$ modulations
in the region with $S_M \ne 0$ and  $S_X \ne 0$,
respectively, but also sometimes  reveal charge redistribution effects 
-- a key point discussed later.

\section{Order parameters}

\label{sec:OP}

To quantitatively characterize the variety of magnetic phases found in the $(N_e, U)$ space, we employ a set of four tailored order parameters that probe both local spin configurations and global periodicities: \( A_m \) (magnetic alignment factor),  scalar product between spins in a $2\times2$ local subcell, quantifies the average local collinearity between neighboring magnetic moments. It is defined as
\begin{equation}
    A_m=\sum^{N_{\rm zone}}_{i,j} \frac{{\bm{M}_i}\cdot{\bm{M}_j}}{\lVert {\bm{M}_i} \rVert \, \lVert {\bm{M}_j} \rVert},
    \label{Scal_prod}
\end{equation}
where $N_{\rm zone}$ corresponds to sites inside a $2\times2$ subcell in the middle of our $6\times6$ supercell; \( T_m \) (magnetic twist vector), the cross product of four neighboring spins, is a sensitive probe to noncoplanarity among groups of spins, capturing vector chirality and geometric torsion. Defined by
\begin{equation}
    T_m= \frac{\left\lVert({\bm{M}_0}\times{\bm{M}_1})\times(\bm{M}_2\times{\bm{M}_3})\right\lVert}{\lVert {\bm{M}_0} \rVert \, \lVert {\bm{M}_1} \rVert \, \lVert {\bm{M}_2} \rVert \, \lVert {\bm{M}_3} \rVert} ,
    \label{Vect_prod}
\end{equation}
it is nonzero only for both the Tetra and Tetra$^*$ configurations. Note also that while local noncoplanar arrangements could in principle yield nonzero signal, such patterns were not observed in our simulations; the structure factors at high-symmetry points \( S_M \) and \( S_X \), corresponding to $2\times2$ and $3\times3$ periodicity respectively, with the $M$ points of the Brillouin zone being linear combinations of $\bm{b}_1/2$ and $\bm{b}_2/2$, $X$-point of $\bm{b}_1/3$ and $\bm{b}_2/3$ $(\bm{b}_1 = \frac{4\pi}{3a}(\frac{\sqrt{3}}{2},-\frac{1}{2})$, $\bm{b}_2 = \frac{4\pi}{3a}(0,1))$, where $a\approx 1.42\,${\rm \AA} is the carbon-carbon distance). They are both defined by
\begin{align}
  S(\bm{q}) & =\sum_{\bm{R}} {\rm e}^{i\bm{q} \cdot \bm{R}} \, \sum_{i,j}^\mathcal{N} \bm{M}_i \cdot \bm{M}_j \,{\rm e}^{i\bm{q} \cdot (\bm{r}_j-\bm{r}_i)},
   \label{Eq_SpinStrucFact}
\end{align}
where $\bm{r}_i$ are the $\mathcal{N}$ site positions in the $L \times L$ supercell.
The sum on the vectors $\bm{R}$ 
of the $L\times L$ supercell switches off $S(\bm{q})$ for any $\bm{q}$ other than 
$\bm{q} = \frac{n_1}{L} \bm{b}_1 + \frac{n_2}{L} \bm{b}_2$.
The evolution of each of these observables is shown in Fig.~\ref{Fig_Order_parameters} for $U = 3t$, computed for both $6\times6$ and $24\times24$ supercells to verify
robustness with respect to the size of the supercell.

For $N_e \le 0.75$, the $24\times 24$ results in Fig.~\ref{Fig_Order_parameters} are very close to the $6\times6$ ones, thus justifying the previous discussion based on the $6\times6$ supercell. In particular, the different order parameters in Fig.~\ref{Fig_Order_parameters} permit us to reconstruct the different phases and the transitions between them along the line $U=3t$ in Fig.~\ref{Fig_Phase_Diag}, with the exception of the stripe phase that is outside the window
$N_e \ge 0.7$ for $U=3t$.
For $N_e > 0.75$, the differences between the $6\times6$ and  $24\times 24$ supercells are larger, indicating that this region requires further attention.

\section{Charge displacement}

\begin{figure}[t!]
\begin{center}
\includegraphics[width=0.8\textwidth]{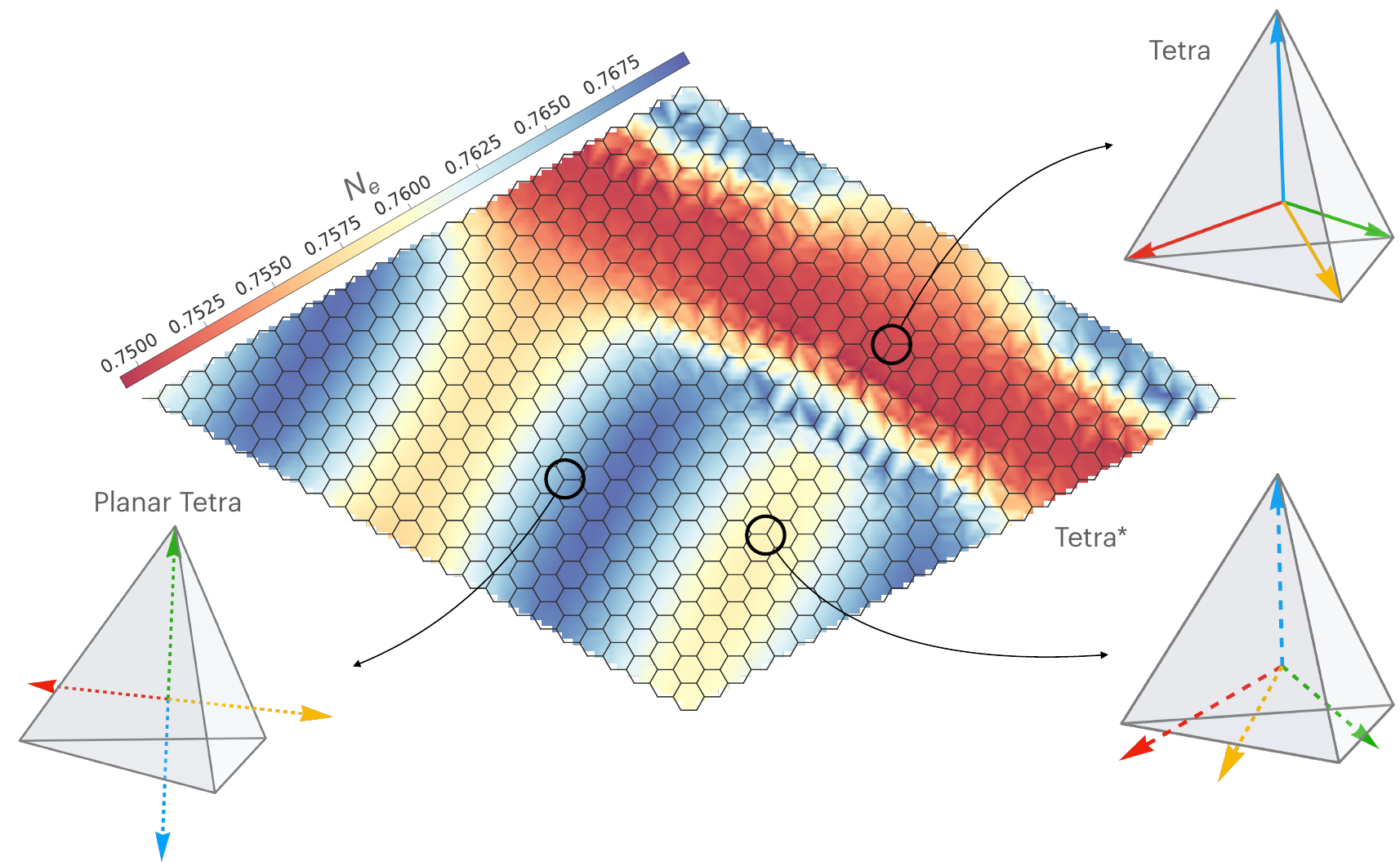}
\caption{\label{Fig_L24_charge_displace}
Charge density in a $24\times24$ supercell at $U=3t$ and average density $N_e=0.76$. We observe non-periodic charge displacement, which creates zones of different density. The red stripe is at $N_e=0.75$ and displays exact tetrahedral order. Blue zones show perpendicular order (4 moments perpendicular to each other, in a plane) with $N_e$ around 0.7675. Yellow zones are at $N_e=0.76$ and show pseudo-tetrahedral orders.
}
\end{center}	
\end{figure}

In the left part of the phase diagram, {\it i.e.},
$0.7 \le N_e \le 0.75$, our mean-field calculations performed in the $6\times6$ and the  $24\times24$ supercells show a small charge displacement between orbitals except for the exact Tetra results, distributed according to the periodicity of the magnetic orders. This displacement is highest for strong interaction value, with a maximum value of $\delta N_e=3 \cdot 10^{-2}$ roughly.

In the right part of the phase diagram, {\it i.e.},
$N_e > 0.75$,
we observed that size frustration (due to $6\times6$ supercell periodicity) may lead to perfect-looking orders which disappear for larger supercells ($24\times24$ supercell). 
However, in the calculations with a larger supercell, a pseudo periodicity of the magnetic order is usually found in 
agreement with  $2\times2$ and $3\times3$ periodicity for the zones $S_M \neq 0$ and $S_X \neq 0$, respectively.

Cases beyond the pseudo $2\times2$ and $3\times3$ periodicity paradigm exhibit a relatively strong charge transfer between different zones of the supercell. 
For instance, Fig.~\ref{Fig_L24_charge_displace} shows that
we lose the global Tetra* order to local orders: in the $24\times24$ supercell, we find Tetra* zones that are incompatible with a $2\times2$ periodicity as neighboring zones break tetrahedric behavior. 
The maximum difference with the average density is small (roughly $\pm0.01$), but essential for the magnetic orders found in each zone. The red regions
in Fig.~\ref{Fig_L24_charge_displace} have an average charge of $N_e = 0.75$ and when looking at the magnetic moments in this region, we find an exact tetrahedral order. Moving away from this zone we lose the tetrahedral angles but keep a 4-moments state, either a Tetra$^*$ order with varying angles or a planar Tetra order. It remains to be checked if these spatial structures are stable for even larger supercells, but our $24\times24$ results underline the importance of the charge density for the magnetic order.
In any case, the spontaneous formation of stripes with charge density $N_e = 0.75$ and accompanying exact tetrahedral order illustrated in Fig.~\ref{Fig_L24_charge_displace}
is remarkably reminiscent of the stripes known from the square-lattice Hubbard model for the high-temperature superconductors
around filling one eighth,
see, e.g., Refs.~\cite{White1998, Service1999, Berg2009, Tranquada2013, Jiang2019, Wietek2022, Hu2024}.
We note that stripes have been observed also in the Hubbard model on the honeycomb lattice, albeit in the lightly doped regime, and with different methods from those employed here \cite{Yang2021}, in particular without admitting for the noncollinear solution that we find here at the mean-field level.

\section{Conclusion}

We have mapped the magnetic phase diagram of doped monolayer graphene near the van Hove singularity using a fully unrestricted, spin-rotationally invariant Hartree-Fock approach to the Hubbard model. This framework allowed us to resolve a rich set of ground states as a function of interaction strength and carrier density, capturing both collinear and complex noncoplanar magnetic textures.

One key result is the stabilization of a fully gapped, noncoplanar tetrahedral magnetic order exactly at quarter doping ($N_e = 0.75$), where the density of states diverges. This state, characterized by zero net magnetization and $2\times2$ periodicity, emerges for any finite $U$ and constitutes a rare realization of a nontrivial spin configuration in a simple one-orbital model. As doping is tuned away from the vHS, a sequence of symmetry-reduced magnetic phases arises, which we have classified using both local spin geometry and structure factor-based order parameters. These include Tetra$^*$, Y and Y$^*$, ferrimagnetic, and stripe states on the underdoped side ($N_e < 0.75$), and increasingly intricate incommensurate textures beyond $N_e > 0.75$, with approximate $2\times2$ or $3\times3$ periodicity.

In the latter regime, we also observe signatures of interaction-induced charge redistribution, reflected in the emergence of spatial density modulations.
These suggest that the electronic correlations near the vHS may drive not only magnetic ordering but also coupling to charge degrees of freedom—possibly a precursor to intertwined or phase-separated states.

The rich magnetic phase diagram revealed here is also reminiscent of phase diagrams of various geometrically frustrated Heisenberg models in an external magnetic field, see, e.g., Refs.~\cite{Zhitomirsky2008,Gvozdikova2011,Gubina2025,Honecker2026} and references therein. However, here this behavior arises as a function of doping, {\it i.e.}, while
conserving the SU(2) symmetry of the model. It is in fact possible to further include a Zeeman term
in the Hartree-Fock treatment of the Hubbard model and it may indeed be interesting to check how the phase diagram is modified and if new phases appear when the symmetry is broken down to U(1) by such an external magnetic field.%

Overall, our results demonstrate that 
graphene doped near the vHS harbors a remarkably rich landscape of magnetic phases, several of which are both nontrivial and robust.
It would be interesting to test the stability of these phases beyond the mean-field approximation.
The density matrix renormalization group method suggests itself for this purpose, as it has been widely
applied to the doped square-lattice Hubbard model, see, e.g., Refs.~\cite{White1998, Jiang2019, Wietek2022, Hu2024}
and references therein.
We are aware of only a few corresponding investigations on the honeycomb lattice carried out so far \cite{Jiang14supercond,Yang2021,Peng2025}, and
given the inherent limitations of the accessible system sizes, we hope that our mean-field results with relatively
large supercells and $k$-space integration will provide useful guidance for the choice of geometries to be used
in future unbiased numerical investigations of graphene doped near $N_e \approx 0.75$.

\section*{Acknowledgements}

Numerical calculations have been performed at {\it Centre de Calculs} (CDC), CY Cergy Paris Universit\'e, 
and at TGCC-GENCI (Project AD010910784). 
We warmly  thank Yann Costes and Baptiste Mary, CDC, for computing assistance. 
We acknowledge financial support from the ANR FlatMoi
project (ANR-21-CE30-0029).

\begin{appendix}\label{appendixA}
\numberwithin{equation}{section}

\begin{figure}[t!]
\centering
\includegraphics[width=0.7\textwidth]{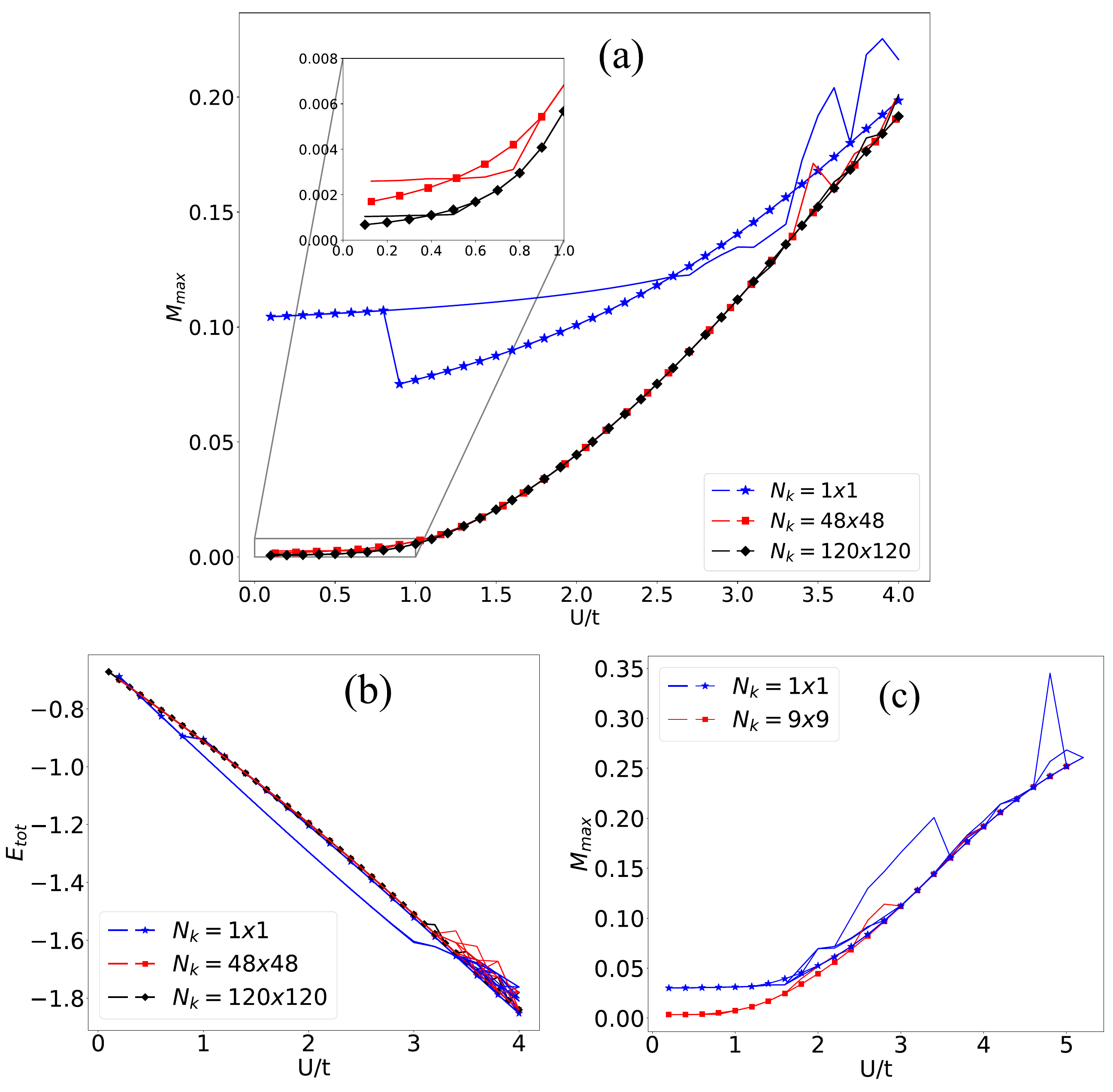}
\caption{\label{Fig_Tetra_U_k}
Maximum local magnetic moment (panels (a) and (c)) and total energy (panel (b)), as a function of the Coulomb interaction $U$ for different values of $N_k$, at $N_e=0.75$. Panels (a) and (b) are for a $6\times6$ supercell, while (c) results from a $24\times2$4 supercell. Lines without markers indicate numerous results using random initial states, while marked lines are for simulations that started from the exact tetrahedral state. 
The inset in panel (a) shows 
the shift between $N_k=48\times48$ and $N_k=120\times120$ at low $U$.
}
\end{figure}

\section{Convergence and importance of the $k$-integration}
\label{appendixC}
This appendix highlights the requirements for sufficiently precise $k$-grids in order to reach the mean-field ground state as well as the tendency to converge towards meta-stable states when starting from a random state.

In Figs.\,\ref{Fig_Tetra_U_k}(a) and \ref{Fig_Tetra_U_k}(c), the maximum local moment for $N_e=0.75$ versus $U$  is compared for different $k$-grids ($N_k$) used for the integration in the reciprocal unit cell for $6\times6$ and $24\times24$ supercells, respectively. 
The corresponding energies for $6\times6$ are shown in Fig.\,\ref{Fig_Tetra_U_k}(b).
The first result is that for large $N_k$, {\it i.e.}, for the most accurate calculation, the Tetra magnetic solution is always the most stable (lowest-energy state).
Moreover, 
we see that for many calculations in the $6\times6$ and $24\times24$ supercells, $N_k=1\times1$ is not sufficient to obtain the real ground state (Tetra): for the $6\times6$ calculations, the lack of $k$-points prevents convergence towards the correct ground state for any value of $U$ even with multiple random initial states or starting from tetrahedral order. As for the larger $24\times24$ supercell at $N_k = 1\times1$, low $U$ values never reach the tetrahedral state but instead a ferromagnetic state, and higher $U$ often converge to a non-collinear and non-tetrahedral meta-stable state. As an example, at $U=2t$, every random initial state computation at $N_k=1\times1$ fails to converge, while the tetrahedral initial state reaches the correct state but at a higher $M_{\max}$ value. Finally, even with high $k$-grid precision, convergence at large $U$ is quite complicated. Indeed, computations for $N_k=120\times120$ sometimes reach the aforementioned meta-stable state for $U>3t$ as shown in Fig.\
\ref{Fig_Tetra_U_k}(a).
This sensitive convergence is also visible at low $U$ where results are different even when $N_k \neq 1\times1$. Indeed $M_{\max}$ is lower (cf.\ inset) for the highest $N_k=120\times120$ even when comparing both tetrahedral initial state computations. We insist on the importance of a precise $k$-grid especially at low interaction values in order to reach ground states and while probing large phase diagram landscapes. 
One last comment about $k$-grid precision lies in the comparison of total energies between states. Integration in $k$-space is an approximation as the correct $k$-grid should be infinite. It follows that the energy values shown in Fig.~\ref{Fig_Tetra_U_k}(b) at $N_k=1\times1$ are not physical. This is understandable when looking at the energies for $N_k=48\times48$ and $N_k=120\times120$ which are very close to each other compared to $N_k=1\times1$, meaning that the $k$-grid is precise enough for valid results and should not vary further with even more $k$-points. Such reasoning is made while looking at the same ground state, here the Tetra one. In order to define which ground state is correct using total energy, it is necessary that the compared results have been computed using the same $k$-grid.

\begin{figure}[t]
\begin{center}
\includegraphics[width=0.6\textwidth]{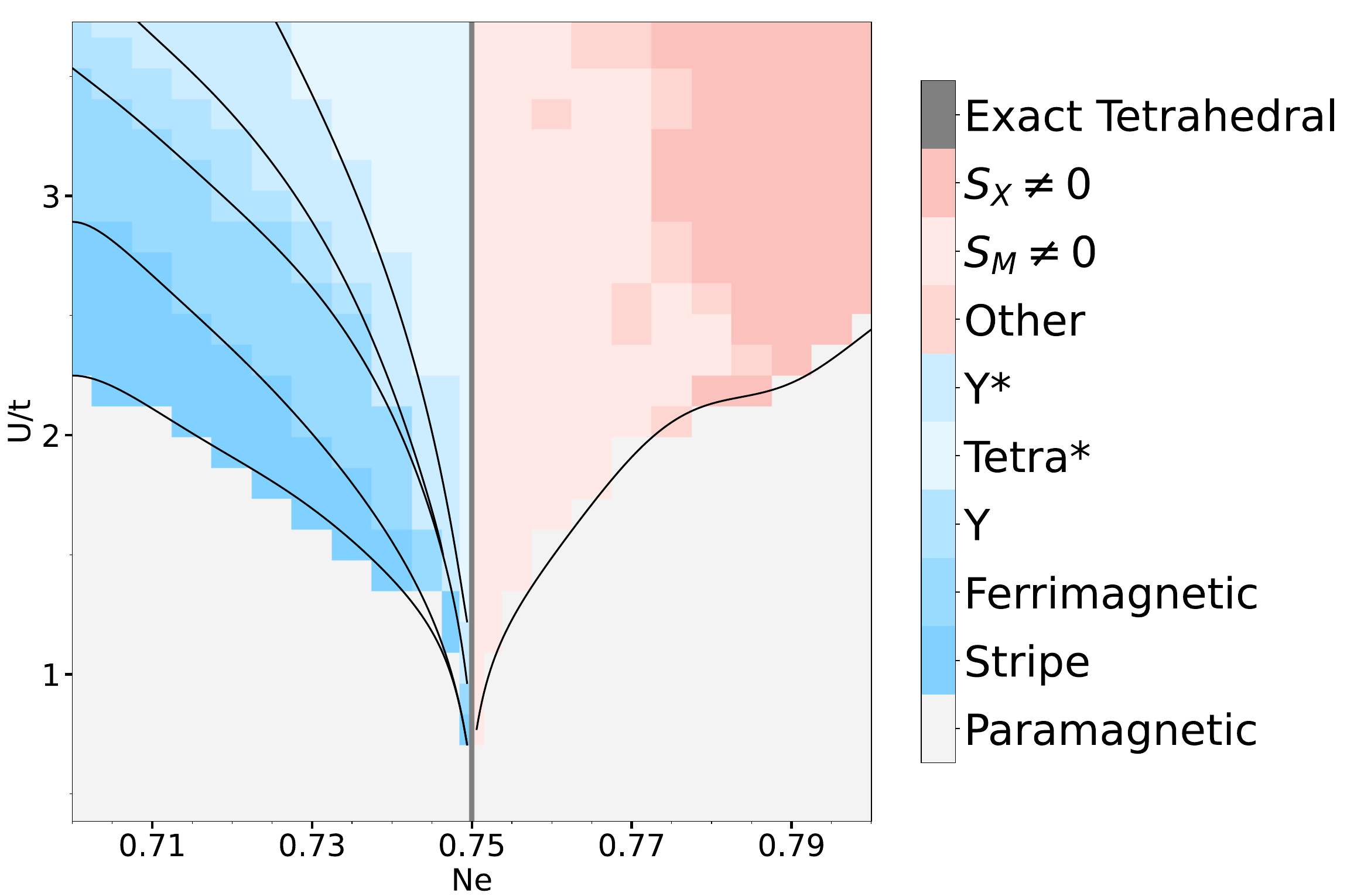}
\end{center}
\caption{\label{Fig_Phase_Diag_raw}
Raw data for the magnetic ground-state phase diagram, with the corresponding transition lines used for the schematic Fig.~\ref{Fig_Phase_Diag}(a). Ground states were selected among 4 sets of results, each with different initial states (Random, Tetra, Y, and Ferri), using lowest total energy per site as the
selection criterion.
}	
\end{figure}

\section{Raw calculated phase diagram}
\label{appendixD}

Figure \ref{Fig_Phase_Diag_raw} shows our calculated raw  data  for the magnetic  ground-state  phase diagram, 
and Fig.\,\ref{fig:2D_proj_6x6} shows
the magnetic configurations in the $6 \times 6$ supercell found at $N_e \le 0.75$. 
To determine the different domains, we used several order parameters (see also Fig.\,\ref{Fig_Order_parameters} and section \ref{sec:OP}).
The scalar product was mainly utilized for the transition between the ferrimagnetic and Y orders using the sharp change in value (we can see the heavy drop around $N_e=0.7175$ in  Fig.\,\ref{Fig_Order_parameters}), although it was also useful to confirm the transition going from Y to pseudo-tetrahedral (Tetra$^*$). We define this magnetic alignment thanks to the average of the spin-spin correlations within a supercell in Eq.\,(\ref{Scal_prod}). Similarly, the vector product was used for detecting the appearance of the Tetra$^*$ order. Indeed, as defined in Eq.\,(\ref{Vect_prod}), this parameter yields zero for the previous order (Y and Y$^*$) while it is non-zero for Tetra$^*$. 
The above order parameters made it possible to accurately determine the boundaries of the domains in the left-hand part of the phase diagram ($N_e \le 0.75$), with exactly the same magnetic order found for the $6\times6$ (Fig.\,\ref{fig:2D_proj_6x6}) and $24\times24$ supercell calculations (see, e.g., Fig.\,\ref{Fig_Order_parameters}).

As explained in the main text, the results obtained on the right-hand side ($N_e > 0.75$) are not identical for calculations in the $6\times6$ and $24\times24$ supercells. 
However, the orders found in the smaller supercell ($6\times6$) are often also found locally in the larger one ($24\times24$).
A criterion common to calculations in both cells is the non-zero value of the spin structure factor (Eq.\,(\ref{Eq_SpinStrucFact})) for specific $\bm{q}$:
$S_M = S(\bm{q}=M)$ and 
$S_X = S(\bm{q}=X)$
corresponding to $2\times2$ and $3\times3$ periodicity (or approximate periodicity), respectively (see Fig.\,\ref{Fig_Phase_Diag}(c)).
\end{appendix}

\bibliography{library}

\end{document}